# Fast Safety Assessment and Correction Framework for Maintenance Work Zones

Zhepu Xu, Qun Yang[*]

*Key Laboratory of Road and Traffic Engineering of Ministry of Education, Tongji University, Shanghai, China, 201804*

[*] Corresponding author, E-mail: qunyang.w@163.com

Acknowledgment: This work was supported by National Nature Science Foundation of China under Grant 51778482.

Word count: 6840 words text.

# Fast Safety Assessment and Correction Framework for Maintenance Work Zones

A framework is proposed to assess the safety of maintenance work zones in a timely manner, show whether there are safety hazards, whether adjustments need to be made and how to adjust it. By means of advanced data acquisition technologies such as multi video detection and portable device based naturalistic driving, the microscopic vehicle behaviour data can be collected. Based on this data, a method for expressing and displaying the distribution of unsafe vehicle behaviour is used to show whether safety hazards exist. Using Vissim, the impacts of the length and speed limit of the warning area, the length and type of the upstream transition area and the length of the work area of the maintenance work zone on the distribution of unsafe vehicle behaviour are simulated to establish the safety correction matrix, which can tell maintenance departments the direction of adjustment when safety hazards exist in maintenance work zones.

## 1 Introduction

Maintenance work zones make the vehicles' operating environment more complicate and decrease road safety. In particular, the road capacity is decreased and the risk of traffic accidents to happen increases. The impact of accidents related to maintenance work zones should not be neglected when planning and executing road maintenance. For example, in the United States, the number of death per year due to accidents in maintenance work zones has laid in the higher hundreds for over 30 years, as it can be seen in Figure 1 (NHTSA 2018). Even though China lacks of an accident database as detailed as the one of the United States, road accidents along maintenance work zones are happening in China too. Considering the ongoing improvements in China's road network, the number of maintenance work and with it maintenance work zones is going to increase in future years. Therefore, the safety of maintenance work zones has to be investigated.

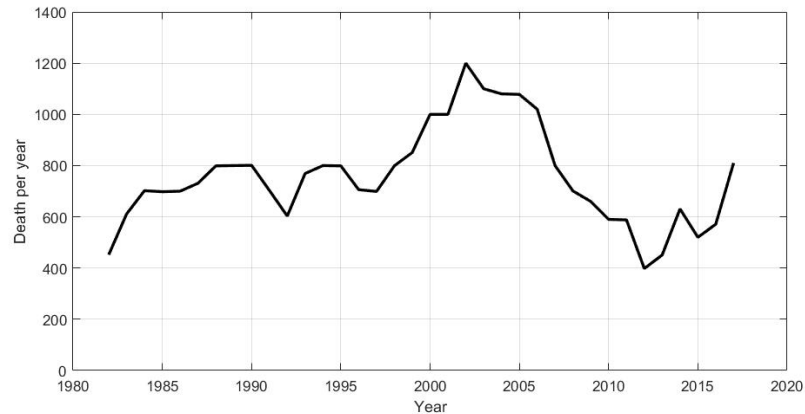

Figure 1 Number of deaths from traffic accidents in maintenance work zones in the US (1982-2017)

In 2004, China issued a first version of Safety Work Rules for Highway Maintenance(Office 2004), which standardized the setting of maintenance work zones. They were improved in a revision 2015 (Transport 2015). This Rules, however, can only propose general guidance and it is impossible to exhaust all the details of a particular maintenance work zone. A common used strategy in practice is to first meet the rules and then make adjustments based on experience. This strategy does not consider safety in an appropriate way and does not allow to make any statement about safety. In fact, even if all the requirements of the rules are met, it is not known how safe the maintenance work zone is, because it may exceed the assumptions of the rules under specific road conditions. For example, over-speed exists often on road sections, which may be neglected in the rules. Whether the ongoing maintenance work zone is safe, what kind of safety hazard exists, and how to correct it is a difficult problem for maintenance departments (Xu and Yang 2018).

Over the past few decades, research on the safety assessment for maintenance work zones has mainly focused on analysing the traffic accident database. The achieved results have been used as guides for the setting of new maintenance work zones and as the main driving force behind the developed rules. However, such approaches can only do post-assessment of maintenance work zones. It is impossible to detect the safety hazards of a current maintenance work zone and to propose corresponding correction methods (Cheng 2006; Weng and Meng 2011). In recent years, with the development of advanced data acquisition technologies, gathering and analysing microscopic vehicle behaviour data (MVBD) has become an interesting tool in order to carry out fast safety assessment for maintenance work zones. For example, Xu and Yang developed a portable MVBD acquisition device, which can collect large sum of MVBD easily. Based on this data, they proposed a method that identifies unsafe behaviour, clusters them in order to find the density centre of the unsafe behaviour, and assesses the safety of the maintenance work zone by analysing the distance from the clustering centre to the maintenance work zone (Xu and Yang 2018). This study, however, assesses only the safety of the buffer area in a maintenance work zone, while other parameters of the work zone cannot be assessed.

This paper is an extension of the above-mentioned research about using advanced data acquisition technologies to gather MVBD to carry out safety assessments of maintenance work zones. A framework is proposed that allows to perform fast safety assessments for entire maintenance work zones, including the warning area, the transition area, the buffer area, the work area, and the termination area, and that enables the suggestion for corresponding correction methods. The framework, data acquisition, safety assessment and the correction process are illustrated on a real world example on an expressway in China. Microscopic traffic simulation is used to be able to analyse the influence of several work zone parameters on the distribution of unsafe vehicle behaviours.

The remainder of this paper is arranged as follows: Chapter 2 consists of a research review including safety assessment methods for maintenance work zones, and advanced data acquisition technologies for collecting MVBD, which is the basis to do fast safety assessment. Chapter 3 presents the fast safety assessment and correction framework. Chapter 4 contains the illustrative example used to identify the impacts of different work zone parameters. The paper is concluded in chapter 5 consisting of the conclusion and remarks for further research.

**2 Related Research**

*2.1 Safety assessment methods for maintenance work zones*

Traditional safety assessment methods for maintenance work zones are mainly based on accident databases. By analysing the factors causing the accidents, knowledge can be gained that guides the design of the layout of future maintenance work zones. As this method derives safety directly from real accident data, it is called direct assessment method (Tang, Zhan, and He 2008). The assessment based on real accident data comes along with some limitations to it. First, this method requires a comprehensive accident database indicating accidents related to work zones separately. Such databases exist only in a few countries, and therefore, there is either no way to use this direct assessment method, or the obtained conclusions are very limited. Second, it allows only for post assessments and cannot be used for timely assessments of new maintenance work zones.

It is in the realization of the limitations of direct assessment method that other surrogate assessment methods are proposed, that describe the relationship between non-accident indicators and safety of work zones. Common non-accident indicators include under others speed difference and traffic conflicts, wherein the traffic conflict technique (TCT) has been widely recognized (Cheng 2006; Weng and Meng 2011; Tang, Zhan, and He 2008). Traffic conflict refers to situations where two vehicles approaching each other and the abnormal traffic behaviour of one of them, such as changing direction, changing speed, sudden braking, etc., would lead to a collision unless the other vehicle performs a corresponding danger-avoiding measure (Haydn 1994). By means of video technology, computer vision or microscopic traffic simulation, the process of collecting vehicle parameters such as the vehicle's trajectory, speed and other data, and performing traffic conflict analysis can be fully automated(Hu 2013; Zhang 2008; Liu 2014). For example, Federal Highway Administration (FHWA) developed the Surrogate Safety Assessment Mode (SSAM) software, which directly imports the results of microscopic traffic simulations into a software for automatic traffic conflict analysis used for safety assessments (FHWA 2018).

Nowadays, the acquisition of MVBD is much easier than before. Researchers can assess the safety of maintenance work zones using different sources, further explore potential risks, and ensure the safety of maintenance work zones. Xu and Yang extracted sharp accelerations out of MVBD to identify unsafe traffic behaviours, since these sharp changes indicate unsafe behaviours. By clustering unsafe behaviour of all vehicles, high-density centres can be identified, which are the most prone position of unsafe behaviour. Considering the spatial relationship between the centres and the maintenance work zones, the safety of the buffer area of maintenance work zones can be assessed (Xu and Yang 2018).

*2.2 Advanced microscopic vehicle behaviour data collection technologies*

Recent advancements in technology allow collecting large amount of MVBD , which is required in order to do fast safety assessments. The two most promising methods for data acquisition are video detection and naturalistic driving.

Video detection based data acquisition is to record the traffic video using a camera, then use computer vision to extract vehicle data from the video, including traffic volume, speed, trajectory and other MVBD . This method is widely used in road network monitoring to serve road safety (Hu 2013; Liu et al. 2012; Zhang 2008). Especially with the development of safety assessment based on traffic conflict technology, it is increasingly becoming an important method to obtain MVBD by video detection (Guo et al. 2016; Autey, Sayed, and Zaki 2012). For example, in (Zhang 2008), cameras were used to record the traffic video at intersections, computer vision method was used to extract the vehicle's trajectory, acceleration and other data, and then traffic conflict analysis was conducted to evaluate the traffic safety based on the data extracted . For the monitoring and data collection of short road sections and intersections, this method is relatively mature. However, the maintenance work zone has its unique characteristics, that is, the maintenance work zone is strip-shaped, and the traditional single-camera method cannot cover the entire maintenance work zone. Fortunately, with the development of video stitching and multi camera detecting technologies, large-scale monitoring problems have been solved (Wang 2013), which brings us new hope to collect the MVBD in maintenance work zones.

Naturalistic driving is to install data acquisition devices on the vehicle, and collect behaviour data of the vehicle or driver in a naturalistic state without disturbing the driver's normal driving (Fitch and Hanowski 2012). This method is often used in driving behaviour related researches, and the scale is increasing (Dingus et al. 2006; Eenink et al. 2014; Regan et al. 2013). Traditional naturalistic driving study installs a large number of sensors in the vehicle. Due to the complicated installation procedure and high installation cost, the vehicles participating in naturalistic driving studies are very limited, especially when comparing with the vehicles running on the road every day, the number of instrumented vehicles, in other word the sample rate is too small, thus, the persuasion of the research results is debatable. In recent years, simplifying naturalistic driving devices, especially using portable devices has become a new trend (Van Ly, Martin, and Trivedi 2013; Johnson and Trivedi 2011; Di Lecce and Calabrese 2008; Dai et al. 2010; Zheng and Hansen 2016). In (Johnson and Trivedi 2011), they used smartphone as the collection device and adopted a crowdsourced model for mass data collection . In (Xu and Yang 2018), portable vehicle behaviour acquisition devices are used, which can collect the MVBD of all vehicles running on the expressway with the aid of highway toll cards for data collection. In particular, China is now vigorously promoting the expressway composite passing cards (Agency 2018), and if such composite cards that can obtain the MVBD are adopted, it will have a very good application prospect.

**3 Fast safety assessment and correction framework**

The fast safety assessment and correction framework put forward in this paper is shown as Figure 2. The framework contains mainly three steps, i.e., (1)collect vehicle behaviour data on site, (2)express and display the distribution of unsafe vehicle behaviour, and (3)make corresponding adjustment according to the safety correction matrix to improve work zone safety, wherein Step (2) has been introduced with detail in the paper (Yang et al. 2019) and can be further divided into three small steps, i.e., identify possible unsafe vehicle behaviour, identify the type of unsafe behaviour and perform spatial analysis to unsafe behaviour. The details of this framework are introduced as follows.

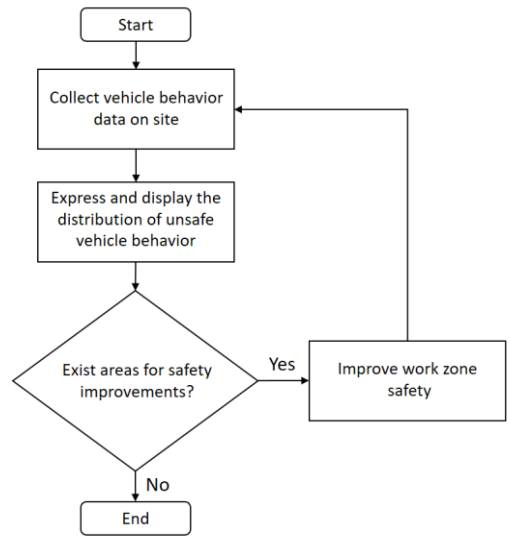

Figure 2 Fast safety assessment and correction framework for maintenance work zones

*3.1 Collect vehicle behaviour data on site*

The first step is to collect the MVBD at the site of the maintenance work zone. In this framework the MVBD is specifically referred to the trajectory, speed, accelerations of vehicle. The behaviour data can be obtained using advanced data acquisition technologies such as video detection or naturalistic driving based data collection methods. An overview of them has been given in the related research chapter before.

*3.2 Identify possible unsafe vehicle behaviours*

After the MVBD of the vehicles are obtained in step 1, unsafe areas are identified by analysing the vehicle behaviour data. A short-time energy based one-parameter bi— thresholds endpoint detection method is applied on both the longitudinal acceleration *ax* and the lateral acceleration *ay* in order to extract unsafe segments from the driving behaviour data. Figure 3, for example, shows in (a) the change of *ax* over time when a vehicle approaches a maintenance work zone, and in (b) the corresponding change of the short-time energy related to *ax*. As can be seen in figure (a), some of the sections fluctuate strongly, which means that the vehicle changes its behaviour, i.e. decelerate. A high short-term energy correlates with an unsafe vehicle behaviour. Using the endpoint detection method, the section with unsafe behaviour can be identified, i.e. the section between the solid green and the dashed red vertical lines in figure (b) can be extracted. The key of the endpoint detection method is the setting of two thresholds T1 and T2, i.e. the solid green and the dashed red horizontal lines in figure (b).

Since the endpoint detection method uses thresholds to identify unsafe vehicle behaviours, the definition of the thresholds is critical. According to previous studies (Chen 2014), the physiology feeling of the driver or passengers to the longitudinal acceleration *ax* and lateral acceleration *ay* is shown in Tables 1.

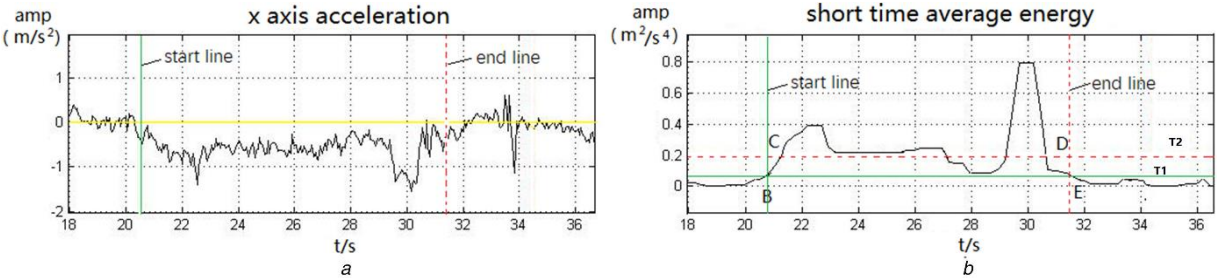

Figure 3 Demonstration of the short-time energy based one-parameter-bi-threshold endpoint detection

Table 1 Physiology feeling of passengers to accelerations

| Type \ Value (m/s²) \ Physiology feeling | Comfortable | General | Uncomfortable |
|---|---|---|---|
| Longitudinal deceleration (absolute value) | ≤1.48 | 1.48~2.46 | ＞2.46 |
| Longitudinal acceleration | ≤0.89 | 0.89~1.25 | ＞1.25 |
| Lateral acceleration | ≤1.8 | 1.8~3.6 | ＞3.6 |

In summary, an unsafe vehicle behaviour is assumed if the lateral acceleration exceeds 3.6 m/s$^2$, the longitudinal acceleration exceeds 1.25 m/s$^2$, or the absolute value of the longitudinal deceleration exceeds 2.5 m/s$^2$. Since the short-time energy is used as the segmentation parameter, the above acceleration thresholds have to be converted into the short-time energy as the value of T2. As for T1, it is mainly for extracting coherent vehicle behaviour segments. After many trials, this study takes the value of the 30% of the short-time energy from large to small as the value of T1.

*3.3 Identify the type of unsafe behaviour*

In a third step, a vehicle behaviour recognition model is established that is based on the support vector machine (SVM) for automatically identifying the behaviour type of an unsafe segment detected in step 2.

In general, 11 types of vehicle behaviour can be identified. These are (1) straight line driving with constant speed (L&C), (2) turning left with constant speed (TL&C), (3) turning right with constant speed (TR&C), (4) turning left and accelerating (TL&A), (5) turning right and accelerating (TR&A), (6) straight line driving and accelerating (L&A), (7) turning left and decelerating (TL&D), (8) turning right and decelerating (TR&D), (9) straight line driving and decelerating (L&D), (10) lane change to the left(TL&CL), and (11) lane change to the right (TR&CL). The training process is consistent with the previous work (Yang et al. 2019), wherefrom the vehicle behaviour recognition can be identified with a 95% accuracy. Therefore, the unsafe behaviour segment identified in step 2 can be accurately identified as one of the 11 types of vehicle behaviour presented before.

*3.4 Perform spatial analysis to unsafe behaviour*

After unsafe behaviours of individual vehicles are detected (step 2) and their types are identified (step 3), the special distribution of all unsafe behaviours are analysed using kernel density analysis. This allows showing the distribution of the different types of vehicle behaviours on a map with their spatial relation to the work zone.

*3.5 Improve work zone safety*

If the kernel density value exceeds a certain threshold, the work zone layout is adjusted according to a safety correction matrix. After the adjustment, the work zone is assessed again (starting with step 1). This is repeated as long as the kernel density value exceeds the threshold.

The safety correction matrix is the core of this framework, which provides criteria for safety assessment, i.e., defines in which case can a work zone be considered as safe or not safe, and gives corresponding correction suggestions when the work zone is not safe. The establishment of the matrix is a process of continuous improvement and requires lots of practice. Chapter 4 takes a real word maintenance work zone as an example to illustrate the process of the matrix establishment.

**4 Example of determining the safety correction matrix**

In the second part, an exemplary study is accomplished, in which an exemplary safety correction matrix is constructed and the influence of several factors of the maintenance work zone on the distribution of unsafe vehicle behaviour is studies. As it is difficult to obtain enough real world data for different layouts of a single work zone, microscopic traffic simulation is used to generate the data set used for the analysis. Vissim is used because it has outstanding performance in simulation fineness, it is widely used in microscopic traffic simulation researches, and it adopts the Wiedemann physiology-psychological model that includes a sophisticated and realistic vehicle following model (Group 2019).

*4.1 Situation*

The example consists of a work zone located on the S20 expressway in Shanghai. As shown in Figure 4, S20 is a two-way eight-lane expressway with a speed limit of 80km/h. The maintenance task is to repair potholes in the two rightmost lanes, and the work zone scheme used is shown in Figure 4. In order to collect vehicle data such as traffic volume and speed, two cameras were placed at the position A and B recording the traffic video. Camera A is located 1 km upstream of the maintenance work zone, where the traffic flow is undisturbed by the work zone there. An OpenCV based software was developed to extract the traffic parameters from the video, i.e. traffic volume and speed distribution of different type of vehicle at the position A and B. This study only distinguishes between small vehicles and large vehicles, wherein small vehicles include cars, small and medium-sized buses and small-sized trucks, and large vehicles include large-sized buses and large and medium-sized trucks. Traffic video data was continuously collected for one hour. The speed distribution at position A and B are shown in Table 2. The traffic volume measured at point A is 1760 vehicles/hour, the proportion of large vehicles is 22%, and for the small vehicles is 78%; the average speed of small vehicles at position B is 75.4km/h, and the average speed of large vehicles is 76.4km/h. The average speed of all vehicles is 75.7km/h.

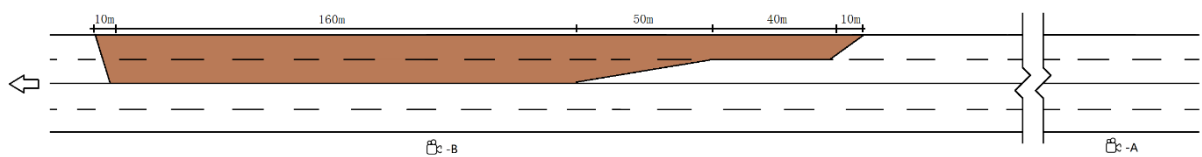

Figure 4 Layout of the measured maintenance work zone

Table 2 Speed distributions in position A and B

|  | Control point(km/h) | 35 | 45 | 55 | 65 | 75 | 85 | 95 | 105 | 110 |
|---|---|---|---|---|---|---|---|---|---|---|
| Position A | Accumulated proportion for small vehicles | 0 | 0.01 | 0.03 | 0.04 | 0.2 | 0.76 | 0.85 | 0.96 | 1 |
| Position A | Accumulated proportion for large vehicles | 0 | 0 | 0 | 0 | 0 | 0.43 | 0.88 | 0.98 | 1 |
| Position B | Accumulated proportion for small vehicles | 0 | 0.01 | 0.09 | 0.28 | 0.65 | 0.86 | 0.97 | 1 | 1 |
| Position B | Accumulated proportion for large vehicles | 0 | 0.02 | 0.07 | 0.30 | 0.63 | 0.84 | 0.93 | 0.98 | 1 |

### *4.2 Vissim calibration*

In order to ensure the feasibility and accuracy of the microscopic simulation results, the simulation model parameters need to be calibrated. There are many parameters that can be set in Vissim, and there are a lot of research on the parameter calibration. Usually the five parameters, i.e., the standstill distance (CC0), headway time (CC1), following variation (CC2), waiting time before diffusion, minimum headway are calibrated (Liu 2012), which are also calibrated in this research. To obtain the optimal set of parameters, a set of orthogonal experiments is designed. The calibration considers 4 levels for each of the 5 parameters and uses a $L_{16}$ ($4^5$) orthogonal table that are shown in Table 3 and Table 4, respectively.

Table 3 Factors and levels considered

| Standstill distance (A) | Headway time (B) | Following variation (C) | Waiting time before diffusion (D) | Minimum headway (E) |
|---|---|---|---|---|
| 0.5 | 0.7 | 3 | 60 | 0.5 |
| 1 | 0.8 | 4 | 80 | 1 |
| 1.5 | 0.9 | 5 | 100 | 1.5 |
| 2 | 1 | 6 | 120 | 2 |

Table 4 Orthogonal table

| Experiment index | Standstill distance (A) | Headway time (B) | Following variation (C) | Waiting time before diffusion (D) | Minimum headway (E) |
|---|---|---|---|---|---|
| 1 | 0.5 | 0.7 | 3 | 60 | 0.5 |
| 2 | 0.5 | 0.8 | 4 | 80 | 1 |
| 3 | 0.5 | 0.9 | 5 | 100 | 1.5 |
| 4 | 0.5 | 1 | 6 | 120 | 2 |
| 5 | 1 | 0.7 | 4 | 100 | 2 |
| 6 | 1 | 0.8 | 3 | 120 | 1.5 |
| 7 | 1 | 0.9 | 6 | 60 | 1 |
| 8 | 1 | 1 | 5 | 80 | 0.5 |
| 9 | 1.5 | 0.7 | 5 | 120 | 1 |
| 10 | 1.5 | 0.8 | 6 | 100 | 0.5 |
| 11 | 1.5 | 0.9 | 3 | 80 | 2 |
| 12 | 1.5 | 1 | 4 | 60 | 1.5 |
| 13 | 2 | 0.7 | 6 | 80 | 1.5 |
| 14 | 2 | 0.8 | 5 | 60 | 2 |
| 15 | 2 | 0.9 | 4 | 120 | 0.5 |
| 16 | 2 | 1 | 3 | 100 | 1 |

For each experiment, the speed distribution, the average speed of the small vehicles, the average speed of the large vehicles, and the average speed of all vehicles at position B are measured and then compared with the actual data. To facilitate the analysis, the speed distribution and the average speeds are integrated into comprehensive indicators according to Formula 1.1 and Formula 1.2, respectively.

$$p1 = \left|\Sigma(a_{sim_{35}} - a_{actual_{35}}) + (a_{sim_{45}} - a_{actual_{45}}) + (a_{sim_{55}} - a_{actual_{55}}) + (a_{sim_{65}} - a_{actual_{65}}) + (a_{sim_{75}} - a_{actual_{75}}) + (a_{sim_{85}} - a_{actual_{85}}) + (a_{sim_{95}} - a_{actual_{95}}) + (a_{sim_{105}} - a_{actual_{105}})\right| \quad (1.1)$$

Where $a_{sim_x}$ and $a_{actual_x}$ are the simulated and actual accumulated proportion of vehicles at the speed control point of x (km/h), respectively.

$$p2 = \left|\Sigma(a_{sim_{small}} - a_{actual_{small}}) + (a_{sim_{large}} - a_{actual_{large}}) + (a_{sim_{all}} - a_{actual_{all}})\right| \quad (1.2)$$

Where $a_{sim_x}$ and $a_{actual_x}$ are the simulated and actual average speed of the x type of vehicles, respectively.

Obviously, for both of these two indicators p1 and p2, the smaller the values, the better the parameters. Figures 5 show for all levels of all parameters the average of all the four experiments including the same level.

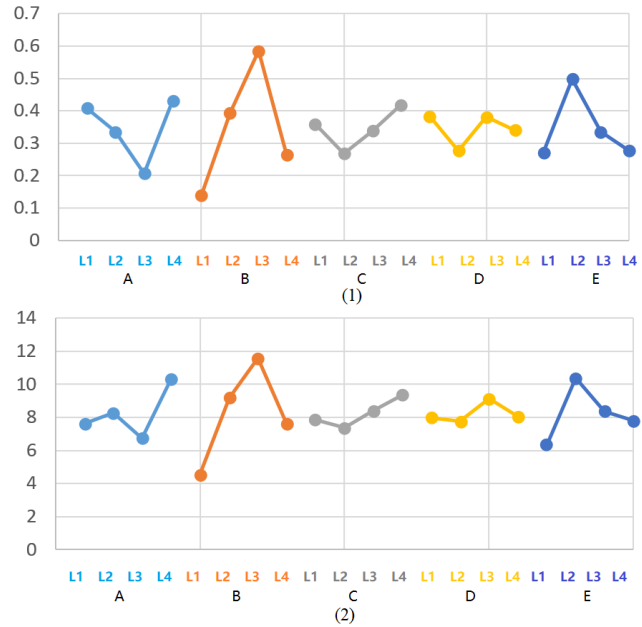

Figure 5  Index values in each experiment, (1)-p1, (2)-p2

Both indicators indicate that $A_3B_1C_2D_2E_1$ is the best solution, i.e., standstill distance(A) taking the value of 1.5, headway time(B) 0.7, following variation(C) 4, waiting time before diffusion(D) 80 and Minimum headway(E) 0.5.

Finally, the best parameters are verified by running another simulation, where the fitness is evaluated according to Formula 2. Comparing the simulation results with the actual data, it is found that the error is within the range of [-10%, 10%] for a confidence of 90%, which means the model is acceptable. Therefore, the above optimization parameters can be used for the simulation analysis later.

$$\xi = \frac{|value_s - value_f|}{value_s} \times 100\% \qquad (2)$$

Where $value_s$ is the actual measured value, $value_f$ is the value obtained by simulation, and ξ is the simulation error.

*4.3 Obtaining vehicle behaviour data in Vissim*

Vissim can output very detailed MVBD with the finest sampling frequency up to 20Hz, i.e. the real-time position, speed and acceleration of the vehicles. This paper uses these MVBD to express the vehicle behaviour. Vissim, however, can only obtain the acceleration in the driving direction, that is, the acceleration in the x direction, and is incapable to obtain the acceleration in the lateral y direction. Therefore, a program is developed to solve the acceleration in two directions according to the position of the vehicle, whose basic principle is illustrated below.

Take (x, y) as the coordinate of the vehicle at time *t*, then the radius of curvature ρ of the trajectory at time *t* is:

$$\rho = \frac{(x'^2 + y'^2)^{\frac{3}{2}}}{x''y' - x'y''}$$

The speed at time *t*:

$$v = \sqrt{x'^2 + y'^2}$$

The longitudinal acceleration *ax*:

$$ax = a_T = \dot{v} = \frac{x'x'' + y'y''}{\sqrt{x'^2 + y'^2}}$$

The lateral acceleration *ay*:

$$ay = a_N = \frac{v^2}{\rho} = \frac{x''y' - x'y''}{\sqrt{x'^2 + y'^2}}$$

*4.4 Example setup*

As shown in Figure 6, a maintenance work zone control area is composed of a warning area, an upstream transition area, a buffer area, a work area, a downstream transition area and a termination area(Transport 2015; MUTCD 2006). Previous paper (Xu and Yang 2018) on safety assessment has mainly focused on the buffer area and neglected the influence of the other areas of a work zone. This work includes the length of the warning area, the speed limit of the warning area, the length and type of the upstream transition area, and the length of the work area.

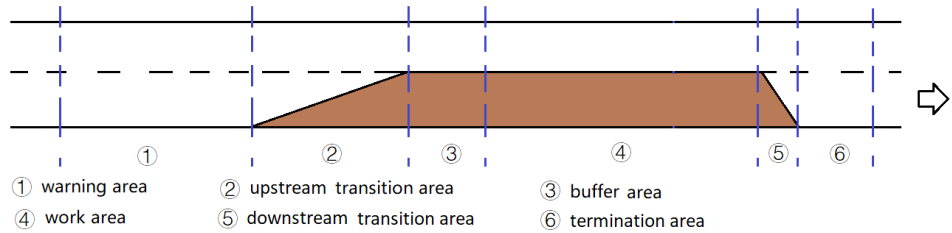

Figure 6 Demonstration of the composition of a maintenance work zone

11 sets of experiments were designed while keeping the driving behaviour parameters unchanged, as shown in Table 5. Scenario 2 is the basic scenario, whose layout is the same as the one shown in Figure 4. Only one parameter value was changed for each other set of experiment, and three parallel experiments were performed for each set of experiment. By adjusting the random seed in Vissim, the vehicle generation in each experiment is different, and the average values of three parallel experiments are taken as the final results.

Table 5 11 sets of experiments

| Factors | Index | Levels | Scenario | Note |
|---|---|---|---|---|
| Length of the warning area | 1 | 300m | Scenario 1 | |
| | 2 | 500m | Scenario 2 | Basic scenario |
| | 3 | 700m | Scenario 3 | |
| Speed limit in the warning area | 1 | 70km/h | Scenario 4 | Only change the speed limit, other factors are the same as Scenario 2. |
| | 2 | 60km/h | Scenario 5 | |
| | 3 | 50km/h | Scenario 6 | |
| | 4 | 40km/h | Scenario 7 | |
| Length and type of the upstream transition area | 1 | Gradual changing-30m | Scenario 8 | Only change the length of the transition area, other factors are the same as scenario 2. The type of the transition area is also included, because Scenario 2 is of stepped style and a gradual changing style is used here. |
| | 2 | Gradual changing-60m | Scenario 9 | |
| | 3 | Gradual changing-90m | Scenario 10 | |
| Length of the work area | 1 | 300m | Scenario 11 | Only change the length of the work area, other factors are the same as Scenario 2. |

It should be noted that this study considers the length of the warning area mainly affecting the distance in which vehicles can change lanes in advance. As shown in Figure 7(1), the vehicles in the 2$^{nd}$ lane cannot change lane to the 3$^{rd}$ lane upstream of the warning area, while they are allowed to change during the warning area.

The speed limit of the warning area can be changed by changing the desired speed distribution in Vissim.

As can be seen in Figure 4, the original transition area has a stepped style. For the example, a gradual changing type is assumed, as illustrated in Figure 7(2).

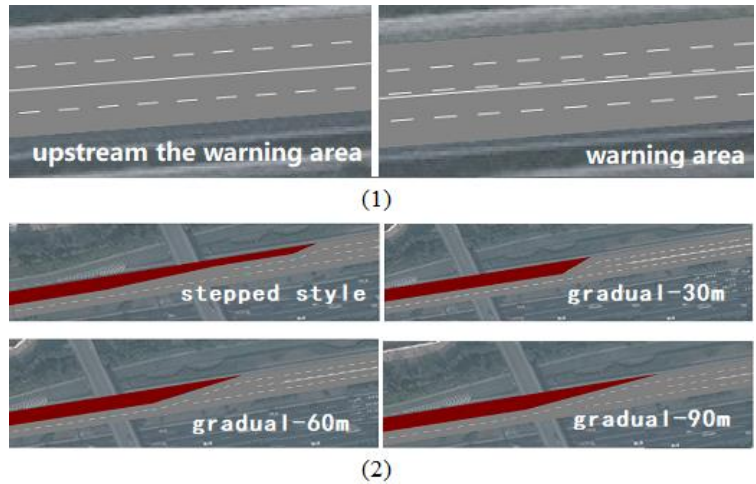

Figure 7 Demonstration of setting of (1) the length setting of the warning area, of (2) the upstream transition area

*4.5 Results*

After each simulation, obtain the MVBD using the method introduced in Section 4.3 and then express and display the vehicle behaviour distribution using the method put forward in paper (Yang et al. 2019). The typical distribution results of the simulation can be seen in Figure 8, where the kernel density distribution maps for each scenario and each existing unsafe behaviour type is shown. The values in the figures are the kernel density extreme values of the vehicle behaviour clustering centres near it, which represent the clustering level of vehicle behaviour distribution, and can be converted into the proportion of vehicles taking the same behaviour according to Formula 3. For example, in Scenario 1 there are three types of unsafe vehicle behaviours existing in the work zone, i.e., L&A, TL&CL and L&D. For TL&CL, the clustering centre shows that most unsafe TL&CLs are distributed in the upstream transition area, and the kernel density value 5.88 shows the maximum percentage of vehicles taking unsafe TL&CL behaviour out of all passing vehicles, which is calculated as 16.1% according to Formula 3, i.e., for every 100 vehicles running through the upstream transition area, almost 16 vehicles will take an unsafe TL&CL behaviour.

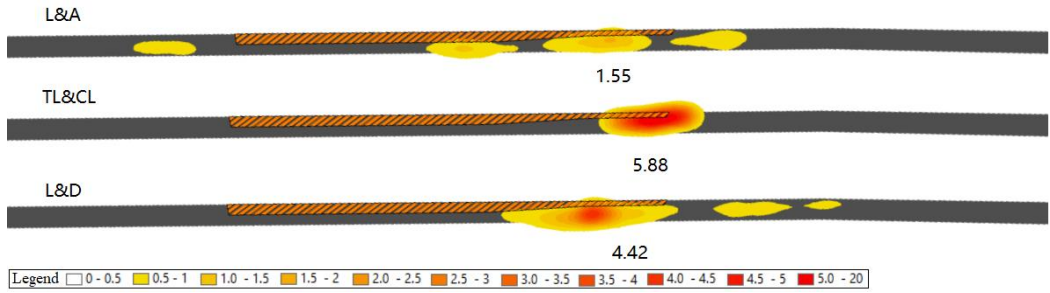

(1) Scenario 1

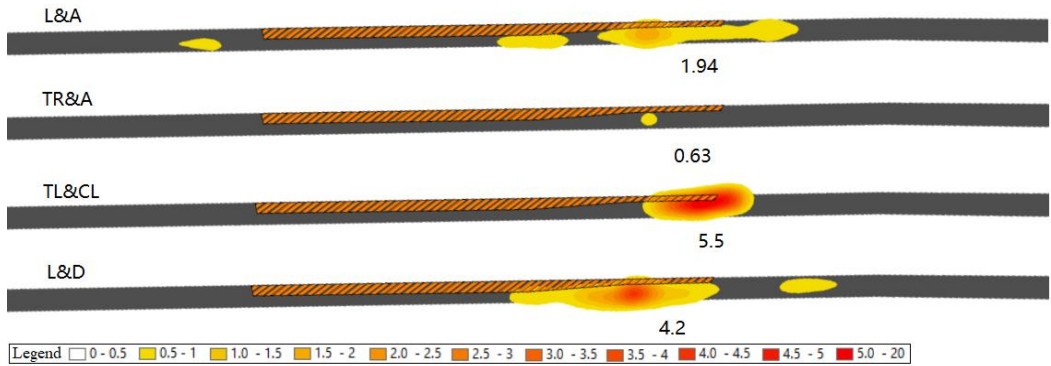

(2) Scenario 2

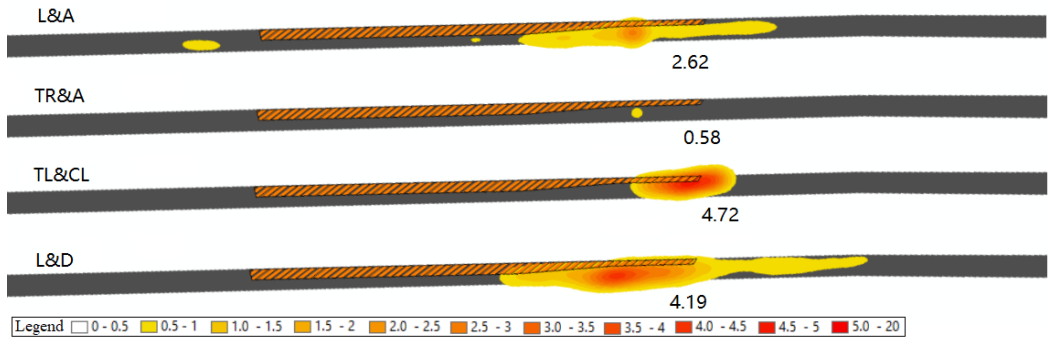

(3) Scenario 3

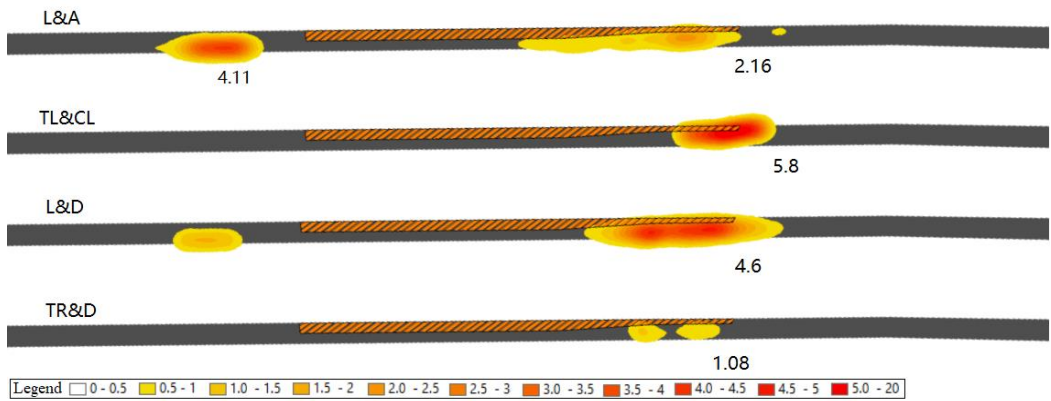

(4) Scenario 4

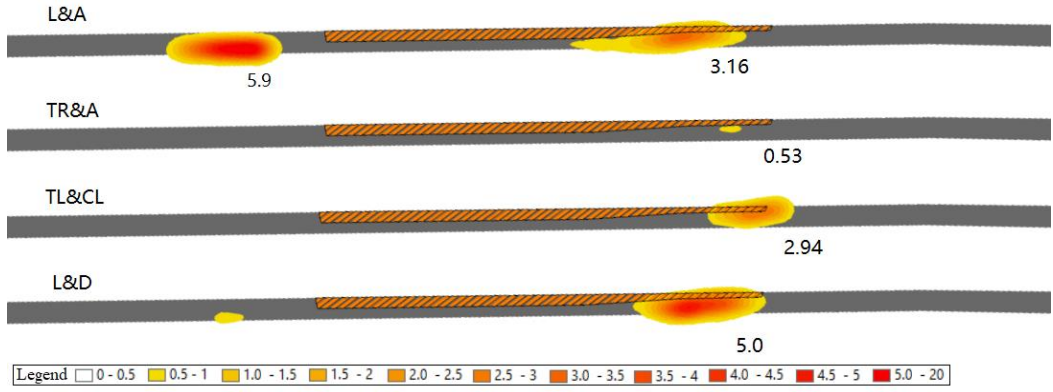

(5) Scenario 5

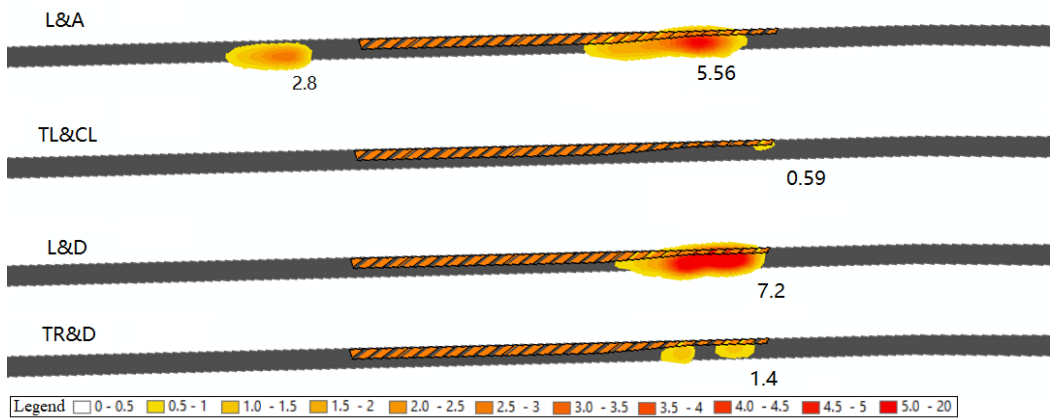

(6) Scenario 6

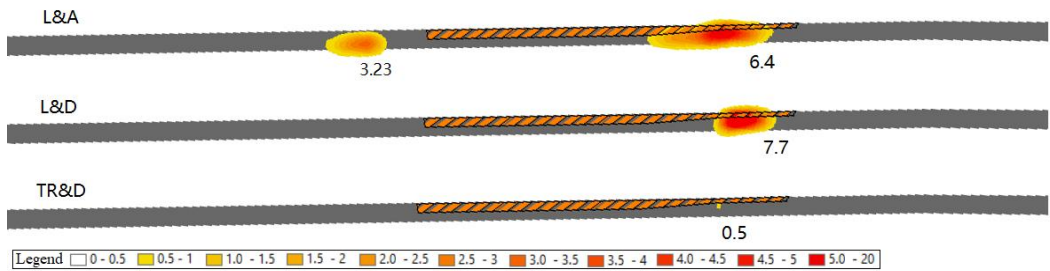

(7) Scenario 7

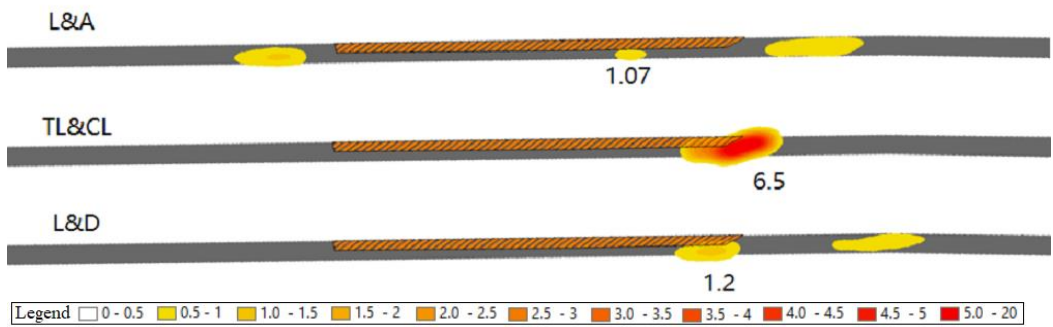

(8) Scenario 8

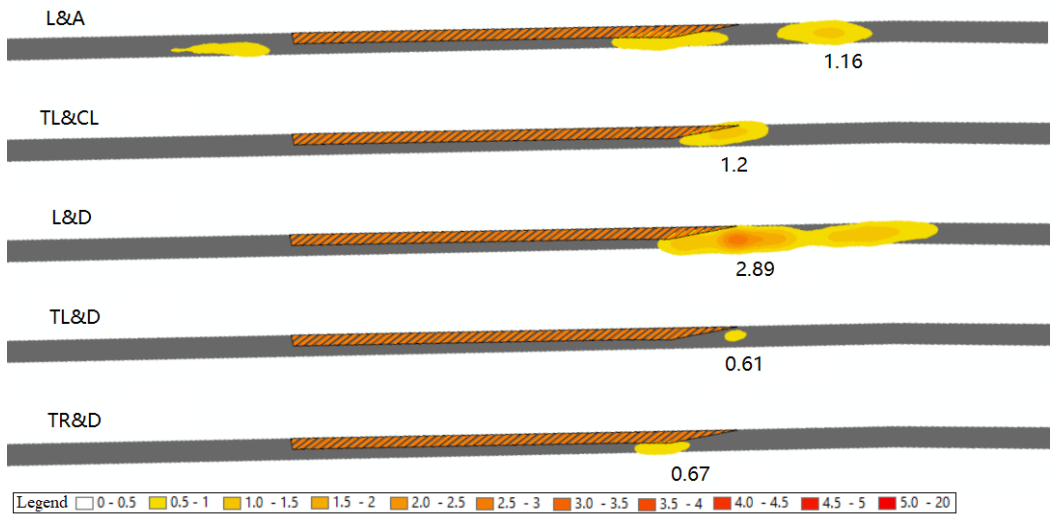

(9) Scenario 9

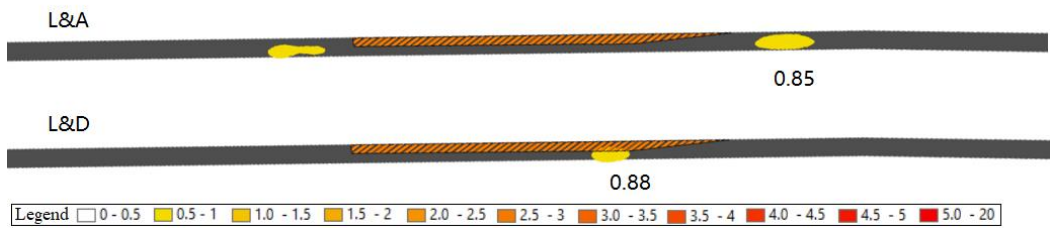

(10) Scenario 10

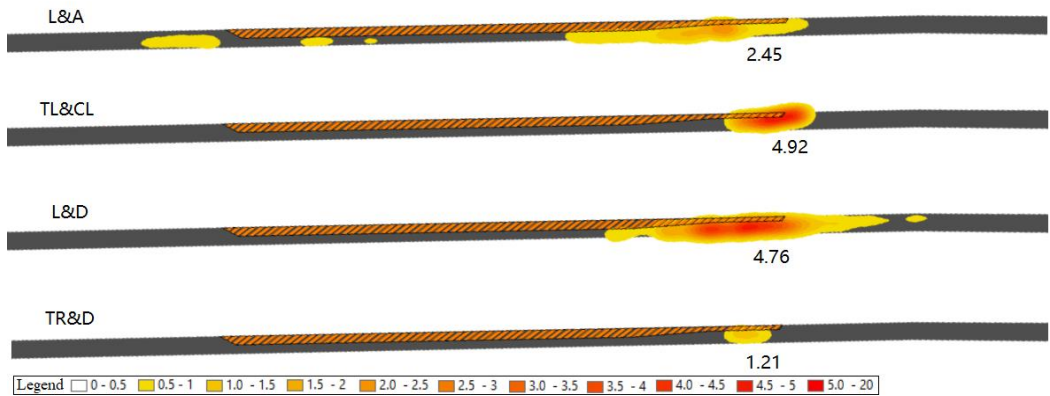

(11) Scenario 11

Figure 8 Typical behaviour distribution maps of the 11 sets of experiments

$$proportion = \frac{kernrl\ density}{36.5} \times 100\% \qquad (3)$$

According to different maintenance work zone parameters, the behaviour distribution types and distribution kernel density extreme values appearing in different scenarios are organized as shown in Table 6. As can be seen from Figure 8, the unsafe vehicle behaviours are mainly concentrated in the upstream part of the work area, and only when the speed limit in the warning area is changed, there will be more unsafe L&A behaviour accumulated in the termination area. Table 6 lists the L&A behaviour distribution in the termination area separately, and the other columns refer to the extreme values of the unsafe behaviour distribution upstream the work area. What should be noted is that the values in the table are the averages of three parallel experiments.

Table 6 Kernel density extreme values in the 11 sets of experiments

| Factors of maintenance work zone | Vehicle behaviour | L&A | TR&A | L&D | TL&D | TR&D | TL&CL | L&A (termination area) |
|---|---|---|---|---|---|---|---|---|
| Length of the warning area | 300m | 1.83 | | 4.36 | | 0.80 | 5.48 | |
| | 500m | 2.57 | 0.63 | 4.73 | | 0.73 | 5.25 | |
| | 700m | 2.72 | 0.58 | 4.73 | | 0.63 | 5.08 | |
| Speed limit of the warning area (km/h) | No speed limit | 2.57 | 0.63 | 4.73 | | 0.73 | 5.25 | |
| | 70km/h | 2.50 | | 4.96 | | 0.85 | 5.92 | 5.23 |
| | 60km/h | 3.13 | 0.53 | 5.58 | | 1.04 | 2.83 | 6.76 |
| | 50km/h | 5.35 | | 7.63 | | 1.02 | 0.59 | 2.42 |
| | 40km/h | 6.78 | | 8.14 | 0.71 | 0.75 | | 3.42 |
| Length or style of the transition area | Gradual-30m | 0.84 | | 1.42 | | | 6.13 | |
| | Gradual-60m | 1.21 | | 2.76 | 0.57 | 0.63 | 1.03 | |
| | Gradual-90m | 0.82 | | 0.85 | | | | |
| | Stepped style | 2.57 | 0.63 | 4.73 | | 0.73 | 5.25 | |
| Length of the work area | 170m | 2.57 | 0.63 | 4.73 | | 0.73 | 5.25 | |
| | 300m | 1.94 | 0.63 | 4.20 | | | 5.50 | |

*4.6 Safety correction matrix*

Considering the results shown in Figure 8 and table 6, the following conclusions can be drawn.

(1) Unsafe vehicle behaviour mainly occurs in the upstream sections of the work zone, especially in the upstream transition area, which is a section where unsafe vehicle behaviours are seriously concentrated and should be given special attention. With a reduced speed limit in the warning area of the work zone, more unsafe behaviours show up in the termination area, where the vehicles speed up again.

(2) The length of the warning area has little effect on the distribution of unsafe vehicle behaviour. Even if the warning area is extended, vehicles will not change lanes early, but take action until the driver notices the maintenance work zone.

(3) The speed limit of the warning area will affect the distribution of the L&A, L&D and lane change behaviours, i.e. the lower the speed limit, the more unsafe L&A and L&D behaviours in the upstream sections of the work area, and the less unsafe lane change behaviour. The speed limit in the warning area will also affect the distribution of unsafe L&A behaviour in the termination area, where the unsafe L&A behaviour first increases and later decrease with the difference between the speed limit in the warning area and the one in normal sections.

(4) The length of the upstream transition area has a great influence on the vehicle lane change behaviour. The longer the length, the more smoothly the transition area changes, and the less unsafe lane change behaviours exist. The stepped style will cause serious traffic disturbances in the transition area and increase the distribution of unsafe acceleration and deceleration behaviours.

(5) The length of the work area has little effect on the distribution of unsafe vehicle behaviour.

Considering the results and the conclusions out of the results, a safety correction matrix for work zones can be obtained, as shown in Table 7.

Table 7 Safety correction matrix

| Index | Problem description | Correction method |
|---|---|---|
| 1 | The kernel density value of the unsafe L&A and L&D behaviours in the upstream of the work area is too large | Appropriately increase the speed limit value of the warning area, and adjust 10km/h each time. If a stepped style transition area is used, change it to a gradual changing style. |
| 2 | The kernel density value of the unsafe lane change behaviour in the upstream of the work area is too large | Increasing the length of the transition area to make the transition area change more smoothly, and adjust 30m each time. In the case where the transition area of the maintenance work zone cannot be adjusted anymore, try to reduce the speed limit of the warning area appropriately. |
| 3 | The kernel density value of the unsafe L&D behaviour in the termination area is too large | Try to appropriately reduce more the speed limit value based on the current speed limit, and adjust the speed by 10 km/h each time. |

*4.7 Discussion*

As shown in Figure 8, the distribution of different vehicle behaviours, including distribution proportions and locations can be intuitively represented on maps, where the kernel density value can be used to characterize the proportion of the distribution. It is, however, difficult to define which kernel density value of the behaviour distribution has to be considered as unsafe and after which reduction a work zone can be considered as safe. The assessment and correction framework presented in this paper allows to increase safety at work zones by adapting correction measures according to the correction matrix. Regarding the identification of safe and unsafe work zones it allows first to analyse the average safety level in terms of the kernel density extreme value of multiple maintenance work zones of a single department. For example, if the maintenance work zone set by a maintenance department causes an average kernel density of the L&A behaviour of 3.1 at the transition area, then a kernel density value higher than 3.1 for the L&A behaviour in the transition area of a new maintenance work zone could be considered unsafe. Second, multiple departments can be compared with each other in terms of their average safety level. Take again the transition area as an example, if the average kernel density value of the L&A behaviour of all departments is 3.1, then a value higher than 3.1 for a specific maintenance work zone of department A can be considered unsafe. Finally, the kernel density value can be related with accident rates based on accident databases, which could be used to determine the threshold of kernel density value for safe and unsafe behaviour.

After establishing the safety correction matrix, the fast safety assessment and correction framework for maintenance work zone is complete. For a new work zone, follow the steps introduced in Chapter 3, then whether the work zone is safe or not can be assessed timely. If not safe, take the corresponding measure in the safety correction matrix and repeat the steps until the work zone is safe. It seems that the fast safety assessment and correction framework proposed in this study is a very time-consuming process that requires repeated adjustments and assessments. It is, however, not exactly true. In fact, when the road traffic volume reaches a medium level, only 20 minutes of collecting traffic data is required in order to perform a reliable assessment. Identifying unsafe areas and taking decisions on the adjustment of the work one is a fast forward process. The number of required iterations in order to identify the best work zone layout is getting reduced with gained experience. Thus, the method proposed in this study is efficient in practical use.

**5 Summary and prospect**

This study proposes a framework to fast assess and correct the safety of maintenance work zones. Using advanced data acquisition technologies such as video detection or naturalistic driving, a large sum of MVBD can be collected. Based on the behaviour data, the distribution of unsafe behaviours in maintenance work zones can be expressed and shown on maps, which intuitively shows the safety hazards of the ongoing maintenance work zone. If necessary, the corresponding adjustment method can be easily determined from the safety correction matrix to eliminate the safety hazards. Wherein, the safety correction decision matrix was obtained using Vissim microscopic traffic simulation to analyse the effects of the length and speed limit of the warning area, the length and type of the upstream transition area and the length of the work area on vehicle behaviour distribution.

There are still other factors of maintenance work zones affecting the distribution of the unsafe vehicle behaviours, which are not all exhausted in this research, but the method put forward in this paper can be used to study the effects of other factors on vehicle behaviour distribution, and constantly improve the safety correction matrix. The next step is to apply the framework to the actual safety assessment for maintenance work zones, establish a database of kernel density values that can reflect the average level of maintenance departments, and compare the kernel density value with the historical traffic accident database to determine the safety level division thresholds.